# Reversibly switchable electromagnetic device with leakage-free electrolyte

Takayoshi Katase[1,*], Yuki Suzuki[2], and Hiromichi Ohta[1,*]

[1]Research Institute for Electronic Science, Hokkaido University, N20W10, Kita, Sapporo 001−0020, Japan

[2]Graduate School of Information Science and Technology, Hokkaido University, N14W19, Kita, Sapporo 060−0814, Japan

[*]E-mail: katase@es.hokudai.ac.jp, hiromichi.ohta@es.hokudai.ac.jp

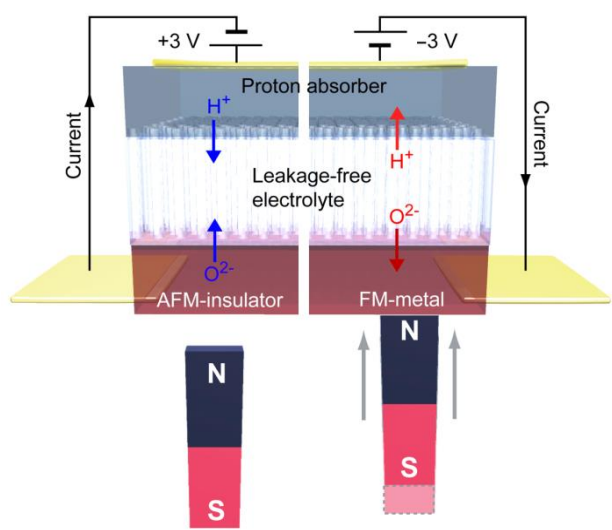

**A reversibly switchable electromagnetic oxide device** is demonstrated by using a thin-film transistor structure with a newly developed "leakage-free electrolyte" as a gate insulator. The electrical and magnetic behavior of the device can be switched from antiferromagnetic insulation to ferromagnetic metal electrically under DC voltage of ±3 V at room temperature. The result provides a novel design concept for practical memory device.

Electromagnetic devices, whose electrical conductance (insulator/metal) and magnetism (non-magnet/magnet) can be switched simultaneously, are expected to become indispensable building blocks for future multiplex writing/reading of electric/magnetic signals. Electrostatic charge-carrier modulation of ferromagnetic semiconductors or metals is considered to be one of the solutions for realizing such electromagnetic devices.[1, 2] However, the aforementioned reversible switching is very difficult because of the high charge carrier concentration. Although the creation of an electric double layer transistor (EDLT) structure, in which the large capacitance of a polarized ionic liquid can be used for charge accumulation/depletion, is a powerful strategy for modulating carrier concentrations,[3, 4] such an approach is not practically applicable because of the liquid leakage problem.

Transition-metal oxides are more preferable as the active material for such electromagnetic devices because of their many switchable functional properties such as electrical conductivity, magnetism, optical absorption, and superconductivity.[5−7] Such switchable features originate from the flexible valence state of the transition-metal ion, which is directly related to the $O^{2-}$ coordination environment of the ion.[8−10] Thus, the electrical conductance and magnetism of transition-metal oxides can be switched simultaneously from the low-conductivity (insulating)/non-magnet (paramagnetic or antiferromagnetic) state to the high-conductivity (metallic)/magnet (ferromagnetic or ferrimagnetic) state by controlling their oxygen off-stoichiometry (**Fig. 1**).

The following two methods are classically known to control the oxygen off-stoichiometry of transition-metal oxides. One is high-temperature heating under an oxidative/reductive atmosphere, because the oxygen off-stoichiometry reaches the chemical equilibrium state within a short time at high temperatures owing to the large diffusion constant of $O^{2-}$ at high temperatures.[11−13] The other is the use of an electrochemical redox reaction in a liquid electrolyte containing an alkali hydroxide as a salt at room temperature (RT).[14] The former method is not preferable for devices operating at RT since high-temperature heating (>500 °C) is commonly required. Although the latter method would be reasonable for RT applications, the device cannot be used without sealing due to the leakage of the liquid electrolyte. Thus, a leakage-free liquid electrolyte is essential for realizing a transition-metal oxide based electromagnetic device.

To overcome this issue, we developed a 'leakage-free electrolyte', incorporated within a nanoporous insulator (**Fig. 2a**). No leakage of the electrolyte in the nanopores occurs because of the large surface tension of the electrolyte/nanopores. First, an alkali-ion containing oxide (insulator) thin film with a nanoporous structure is fabricated on a transition-metal oxide layer. The water vapour from air is expected to be automatically absorbed into the nanopores due to capillary action;[15, 16] then, a small amount of alkali ions from the alkali-ion containing oxide is dissolved into the nanopores, and finally, the nanopores are filled with the alkali hydroxide aqueous solution, which can be used as an electrolyte for the electrochemical redox reaction of transition-metal oxides, *e.g.* manganese oxide in alkaline batteries.[17] Subsequently, a proton absorber layer is deposited on the nanoporous alkali oxide film to supress $H_2$ generation during the device operation.

We focused on $NaTaO_3$, a promising photocatalyst for the generation of $H_2$ gas from water,[18] with the assumption that a small amount of $Na^+$ ions from $NaTaO_3$ can be dissolved in water. We finally found that the $NaTaO_3$ film, which can be deposited by pulsed laser deposition under a relatively high oxygen atmosphere (11 Pa) at RT, satisfies the requirement. The $NaTaO_3$ film is a nanoporous glassy oxide with a nanopillar (10−20 nm in diameter) array structure (**Fig. 2b**). The film density is 3.3 g cm$^{-3}$, which is ~47 % of fully dense $NaTaO_3$ film (7.0 g cm$^{-3}$) deposited under low oxygen atmosphere (< 5 Pa), indicating that the film has a large amount of space to absorb water vapour from the air.

We then measured the conductivity of the $NaTaO_3$ nanopillar array film, sandwiched by Au electrodes. **Figure 2c** shows a representative Cole-Cole plot of the $NaTaO_3$ nanopillar array film obtained at RT, in air (amplitude = ±10 mV). A semicircle is clearly seen in the high-frequency region due to mobile charge. The conductivity was calculated to be 2.5 μS cm$^{-1}$, three orders of magnitude larger than that (5 nS cm$^{-1}$) of the dense $NaTaO_3$ film and two orders of magnitude larger than that (55 nS cm$^{-1}$) of ultra-pure water.[19] In order to clarify the contribution of water to the conductivity, we also obtained a Cole-Cole plot of the $NaTaO_3$ nanopillar array film in vacuum (**Fig. 2d**). The calculated conductivity was 34 pS cm$^{-1}$, which was five orders of magnitude smaller than that in air. Further, we observed water electrolysis ($H_2$ /$O_2$ gas generation, **Movie S1**), when measuring the impedance of the $NaTaO_3$ nanopillar array film at a low frequency of 5 Hz, with an amplitude of ±3 V. From these results, we judged that water vapour in the air is incorporated into the $NaTaO_3$ nanopillar array film, and then, a small amount of $Na^+$ ions from the $NaTaO_3$ nanopillars is dissolved into the nanopores; finally, the nanopores are filled with aqueous NaOH solution, which serves as a leakage-free electrolyte.

To demonstrate the potential of the leakage-free electrolyte for an electromagnetic device, we selected $SrCoO_x$ as the transition-metal oxide,[20, 21] primarily because the reversible redox reaction of $SrCoO_x$ proceeded at RT with an aqueous solution of an alkali hydroxide as the liquid electrolyte.[22–25] Further, the electromagnetic phase of this oxide can be switched from antiferromagnet (AFM; non-magnet)-insulator to ferromagnet (FM; magnet)-metal by controlling the oxygen off-stoichiometry ($x$) from 2.5 to 3.0. The $SrCoO_{2.5}$, with an oxygen-vacancy-ordered brownmillerite structure, is an AFM insulator with trivalent $Co^{3+}$, whereas $SrCoO_3$, with a simple perovskite structure, is a FM metal with tetravalent $Co^{4+}$.[26–28] We expected that under a negative gate current, the $x$-value of $SrCoO_x$ switches from 2.5 to 3.0 due to oxidation by $OH^-$ ions. In contrast, under a positive gate current, the $x$-value reverts to the virgin state, thus resulting in reversible AFM-insulator/FM-metal switching.

Here, we demonstrate a RT-operable electromagnetic device based on a transition-metal oxide, $SrCoO_x$, using the leakage-free electrolyte incorporated in the amorphous $NaTaO_3$ nanopillar array film. The resultant device showed excellent characteristics: reversible AFM-insulator/FM-metal switching of the transition-metal oxide occurred electrically under a small DC voltage (±3 V) within a very short time (2–3 s) at RT.

We fabricated a transition-metal oxide based electromagnetic device (**Fig. 3a**), whose structure is similar to that of conventional three-terminal thin-film transistors composed of a transition-metal oxide layer (20-nm-thick $SrCoO_{2.5}$) as the active channel layer (**Fig. S1**), the leakage-free electrolyte incorporated in a 200-nm-thick amorphous $NaTaO_3$ nanopillar array film as the gate insulator, and a proton absorber (20-nm-thick amorphous $WO_3$, a well-known electrochromic material [29]) / metal bilayer as the gate electrode. This layer structure was confirmed by transmission electron microscopy (TEM) image (**Fig. 3b**). As the present device structure can be considered as a nano-sized electrochemical cell, in which the $SrCoO_x$ and $WO_3$ layers are respectively used for $OH^-$ and $H^+$ absorber layers, the $H_2$ / $O_2$ gas generation is totally suppressed and it realizes the cycled electrochemical reaction in the solid device. The channel length and channel width were 800 μm and 400 μm, respectively. The device switching was tested at RT in air without sealing.

First, a negative gate bias ($V_g$) of –3 V was applied for 3 s and the electromagnetic properties of SrCoO$_x$ layer were then measured after switching the $V_g$ off, becuase of the non-volatile device operation. The color of SrCoO$_{2.5}$ layer was brown in the virgin state but turned grey after the $-V_g$ application. In the virgin state, the SrCoO$_{2.5}$ layer showed insulating behaviour (**Fig. 3c, A**); an exponential increase in sheet resistance ($R_s$) was observed as a function of temperature. In addition, the magnetic moment ($m$) was negligibly small, with no magnetic transition in the measured temperature range (**Fig. 3d, A**), suggesting the AFM nature of SrCoO$_{2.5}$.[20] It should be noted that it is difficult to judge the AFM state in the virgin SrCoO$_{2.5}$ layer only from the magnetic properties in **Fig. 3d**; considering that (1) it is a known fact that the brownmillerite SrCoO$_{2.5}$ is an AFM insulator with $T_N$ = 570 K,[20] (2) it is confirmed that the SrCoO$_{2.5}$ layer is an insulator with the brownmillerite structure, in which the alternately stacked structure of CoO$_4$ tetrahedra and CoO$_6$ octahedra layers are clearly observed (**Fig. S1**), and (3) the magnetic susceptibility of the SrCoO$_{2.5}$ layer is one order of magnitude smaller than that estimated by Curie's law of paramagnetism [30] and consistent with that previously reported in brownmillerite SrCoO$_{2.5}$ film with AFM state,[31] the virgin SrCoO$_{2.5}$ layer is considered to be AFM state rather than paramagnetic state. Upon applying a $V_g$ of –3 V, $R_s$ decreased by more than three orders of magnitude at RT, and the layer showed metallic-temperature dependence (**Fig. 3c, B**). Furthermore, FM transition was observed at a Curie temperature of 275 K (**Fig. 3d, B**), which originates from the fully oxidized SrCoO$_3$ phase.[32] The saturation magnetization at 10 K (inset of **Fig. 3d**) was ~2.0 $\mu_B$/Co, which is almost the same as that of bulk SrCoO$_3$ (2.1 $\mu_B$/Co).[21] In contrast, upon the application of $V_g$ = +3 V for 3 s, the $R_s$- and $m$-values reverted to the virgin state (**Figs. 3c and 3d, C**). We further tested the cyclability of the electromagnetic device (inset of **Fig. 3c**). Upon the application of ±3 V for 2−3 s at RT in air, reversible switching of $R_s$ (ON/OFF ratio ~10$^3$) was realized.

Crystal structure evolution in the states A–C can be seen in the X-ray diffraction patterns (**Fig. S2**). The diffraction peaks originating from the oxygen vacancy ordered structure ($l$ = 6, 10) were completely suppressed, without destruction of the perovskite structure, clearly confirming the successful oxidation of SrCoO$_{2.5}$ with the brownmillerite structure and the formation of SrCoO$_3$ with the perovskite structure.[32] The $c$-axis lattice constant of the perovskite unit cell steeply decreased from 0.393 nm to 0.379 nm upon the application of $V_g$ = –3 V, maintaining the coherent epitaxial relationship with the SrTiO$_3$ substrate (**Fig. S3**). These results clearly indicate that the entire region in the SrCoO$_x$ layer can be reversibly switched from AFM insulator ($x = 2.5$) to FM metal ($x = 3.0$), electrically under a small DC voltage (±3 V) within a very short time (2–3 s) at RT.

In order to clarify the device operation mechanism, we investigated the relationship between flowing current under various $V_g$ applications (**Fig. 4**). First, we measured the retention-time ($t$) dependence of gate current ($I_g$) along with $R_s$ for the SrCoO$_x$ layer under various ±$V_g$ values ranging from 1.5 to 3.0 V at 1 s intervals, where a negative $V_g$ was initially applied for oxidation (**Fig. 4a**) and a positive $V_g$ was subsequently applied for reduction (**Fig. 4b**). The $I_g$ were first measured during the $V_g$ application and the $R_s$ were then measured after switching the $V_g$ off; the measurement sequence was performed repetitively. Upon negative $V_g$ application (**Fig. 4a**), $-I_g$ increased with $t$ for each $V_g$ and $R_s$ simultaneously decreased, indicating electrochemical oxidation of the SrCoO$_x$ layer. Meanwhile, upon positive $V_g$ application (**Fig. 4b**), $I_g$ decreased with $t$ for each $V_g$ and $R_s$ simultaneously increased due to electrochemical reduction. It should be noted that the initial $R_s$ for positive $V_g$ application (**Fig. 4b**) is same with the final $R_s$ for negative $V_g$ application (**Fig. 4a**), because the conducting state remains unchanged unless the positive $V_g$ is started to be applied.

In both cases (**Figs. 4a and 4b**), the switching time required for the redox reaction largely depends on $V_g$. The time scale of the switching speed can be considered to depend on ionic polarization in the electrolyte, electric double layer formation, electrochemical surface reaction, and O$^{2-}$ ion diffusion into SrCoO$_x$ layer under $V_g$ application. The rate-determining step among them should be the surface reaction and/or O$^{2-}$ ion diffusion in the SrCoO$_x$ film at RT.[33] Therefore, the switching speed should strongly depend on the density of OH$^-$ ions accumulated at the SrCoO$_x$ surface, because the surface exchange reaction is reported to largely depend on the oxygen pressure.[34] This supports the fact that the switching speed depends on the applied $V_g$, as is demonstrated by the exponential increase in the current density in the electrolyte with $V_g$.

**Figure 4c** compares the variation in $R_s$ with electron density, as calculated from the integral value of the $I_g$–$t$ plots in **Figs. 4a and 4b**. $R_s$ decreased and increased along the universal line under all values of $V_g$, and the oxidation and reduction processes were completed at ~7.0×10$^{16}$ cm$^{-2}$, which corresponds well with the ideal value (7.0×10$^{16}$ cm$^{-2}$) required for the redox reaction between SrCoO$_{2.5}$ and SrCoO$_3$, according to the following equation: SrCoO$_{2.5}$ + 2$x$OH$^-$ ⇆ SrCoO$_{2.5+x}$ + $x$H$_2$O + 2$x$e$^-$ (0 ≤ $x$ ≤ 0.5). The universal change in $R_s$, regardless of the applied $V_g$, indicates that all the provided electrons were used for the electrochemical redox reaction of the SrCoO$_x$ layer, obeying Faraday's laws of electrolysis, and that the device operation can be controlled by current density.

The present redox approach involving transition-metal oxides using a leakage-free electrolyte has several merits as compared to classically known methods. Primarily, the device can be operated at RT, although high-temperature heating is required when utilizing simple O$^{2-}$ diffusion in the oxides under an oxidative/reductive atmosphere. Further, the present device can be used for practical applications without sealing, as opposed to known electrochemical devices such as alkaline batteries and electrolytic capacitors, which have the liquid leakage problem. This is an important advantage against recently developed EDLTs,[3,4] which cannot be used without sealing as well. Furthermore, amorphous NaTaO$_3$ is a chemically stable rigid glassy solid, showing good reliability. These clearly indicate that the present approach utilizing a leakage-free electrolyte is epoch-making, providing a novel concept for the design of electromagnetic devices. Since there are many other redox systems in transition-metal oxides, which can be made to exhibit various properties such as electrochromism,[35] thermochromism,[36] and superconducting transition[37, 38] by control of their oxygen off-stoichiometry, the present approach may be useful for the development of not only new electromagnetic devices but also new electro-optical and electrothermal devices.

In summary, we have demonstrated a RT-operable electromagnetic device based on a transition-metal oxide, SrCoO$_x$, using a newly developed leakage-free electrolyte incorporated in an amorphous NaTaO$_3$ nanopillar array film. The resultant device showed excellent characteristics: reversible AFM-insulator/FM-metal ($T_c$ = 275 K) switching of the SrCoO$_x$ layer occurred electrically under a small DC voltage (±3 V) within a very short time (2–3 s) at RT in air, *i.e.* the device is operable with two alkaline batteries (3 V) within the second time scale. The present device does not present liquid leakage problems, and the leakage-free electrolyte provides a novel design concept for transition-metal oxide-based practical electromagnetic devices, whose electrical conductance and magnetism can be simultaneously switched. Such devices are expected to be indispensable building blocks for multiplex writing/reading of electric/magnetic signals. Further, the present approach may aid in the development of many new devices based on transition-metal oxides.

**Experimental**

**Fabrication and characterization of 'leakage-free electrolyte'**: 200-nm-thick amorphous NaTaO$_3$ films were fabricated by pulsed laser deposition (PLD, KrF excimer laser) using stoichiometric NaTaO$_3$ ceramic disk as the target at RT under an oxygen atmosphere (11 Pa for nanopillar array film, <5 Pa for dense film). The thickness and density of the NaTaO$_3$ films were evaluated by X-ray reflectivity measurements. The nanopillar array structure of the NaTaO$_3$ films was observed by using Z-contrast high-angle annular dark-field scanning electron microscopy (HAADF-STEM, JEM-ARM200F, 200 kV, JEOL Ltd.). The ionic conductivity of the NaTaO$_3$ films was measured by the AC impedance method using a YHP4192A impedance analyser and a VersaSTAT 4 Potentiostat/Galvanostat (Princeton Applied Research Ltd.), where the Au/NaTaO$_3$/Au sandwiched structure was fabricated on glass substrate and the AC conductivity was measured along the cross-plane direction at RT.

**Device fabrication**: Thin-film-transistor-type electromagnetic devices were fabricated on 20–40-nm-thick SrCoO$_{2.5}$ epitaxial films using stencil masks.[16] The dimensions of the channel were 400 μm (width) × 800 μm (length) for electrical transport measurements, and 2.0 mm (width) × 2.0 mm (length) for crystallographic characterization and magnetic measurements. The SrCoO$_{2.5}$ film was grown at 720 ºC under an oxygen atmosphere of 10 Pa on a (001) SrTiO$_3$ single crystal substrate (10×10×0.5 mm$^3$) by PLD (1.5 J cm$^{-2}$ pulse$^{-1}$, 10 Hz). The resultant SrCoO$_{2.5}$ films were coherently grown on a SrTiO$_3$ substrate with an epitaxial relationship of (001)[110] SrCoO$_{2.5}$ ∥ (001)[100] SrTiO$_3$. Then, 200-nm-thick NaTaO$_3$ films were deposited under $P_{O2}$ of 11 Pa, and amorphous WO$_3$ films (20 nm) were deposited as the proton absorber under $P_{O2}$ of 10 Pa by PLD at RT. 20-nm-thick Cr/Au bilayer film was used for the source and drain electrodes, and a 50-nm-thick Au or Ti film was used for the gate electrode, where the films were deposited by electron beam evaporation in a vacuum (~10$^{-4}$ Pa, no substrate heating).

**Structural analyses of the devices**: The crystal structures, including the crystallographic orientation and the mosaicity of the SrCoO$_x$ films, were investigated by high-resolution XRD (Cu K$α_1$, ATX-G, Rigaku Co.) at RT. The film surface morphologies were observed by atomic force microscopy (Nanocute, HITACHI High-Tech Ltd.). The cross-sectional microstructure of the resultant electromagnetic device was observed by high-resolution TEM and HAADF-STEM (JEM-ARM200F, 200 kV, JEOL Ltd.), where the electron incident direction was SrTiO$_3$ [110].

**Electromagnetic property measurements of the devices**: $I_g$ during $V_g$ application was measured by using a source measure unit (Keithley 2450). $R_s$ was measured by the DC four probe method in the van der Pauw configuration. The DC magnetization was measured in the temperature range of 10−300 K using a magnetic property measurement system (MPMS, Quantum Design Ltd.). The electromagnetic properties of SrCoO$_x$ layer were measured after applying $V_g$ and switching it off becuase of the non-volatile device operation.

**Acknowledgements**
We thank M. Yamanouchi for discussion and N. Kawai for the TEM/STEM analyses. The TEM/STEM analyses, conducted at Hokkaido Univ., were supported by Nanotechnology Platform Program from MEXT. T.K. was supported by Grant-in-Aid for Young Scientists A (15H05543) from JSPS. H.O. was supported by Grant-in-Aid for Scientific Research A (No. 25246023), Grant-in-Aid for Scientific Research on Innovative Areas (No. 25106007) from JSPS.

**Figure 1.** Concept of an electrically switchable electromagnetic device. Using the flexible value of the valence state of transition-metal ions in transition metal oxides, the device can be switched from an insulating/non-magnet (paramagnetic or antiferromagnetic) state to a metallic/magnet (ferromagnetic or ferrimagnetic) state simultaneously by electrochemical oxidation / reduction (redox) reaction at room temperature in air. Such a device would be an indispensable building block for multiplex writing/reading of electric/magnetic signals.

**Figure 2.** Leakage-free electrolyte. (**a**) Schematic device structure composed of transition-metal oxide, leakage-free electrolyte, and proton absorber layer. (**b**) Cross-sectional HAADF-STEM image of a 200-nm-thick $NaTaO_3$ film deposited on $SiO_2$/Si substrate. A nanopillar array structure is seen. A broad halo is observed in the electro-diffraction pattern. (**c**) Cole-Cole plot of a 200-nm-thick $NaTaO_3$ nanopillar film (0.95 mm × 0.80 mm) measured at RT in air. The conductivity was 2.5 µS cm$^{-1}$, three orders of magnitude larger than that (5 nS cm$^{-1}$) of a dense $NaTaO_3$ film in air and two orders of magnitude larger than that (55 nS cm$^{-1}$) of ultra-pure water. (**d**) Cole-Cole plot of the film measured in vacuum (~10$^{-2}$ Pa). The calculated conductivity was 34 pS cm$^{-1}$, five orders of magnitude smaller than that in air. First, water vapour in air atmosphere is incorporated in the $NaTaO_3$ nanopillar array film, then small amount of Na$^+$ ions of the $NaTaO_3$ nanopillars is dissolved into the nanopores, and finally, the nanopores are filled with NaOH aqueous solution, which serves as a 'leakage-free alkali electrolyte'.

**Figure 3.** Demonstration of an electrically switchable electromagnetic device. (**a**) Schematic device structure, which is similar to conventional three-terminal thin-film transistors, composed of a 20-nm-thick $SrCoO_{2.5}$ active channel layer, 'leakage-free alkali electrolyte' (200-nm-thick amorphous $NaTaO_3$ nanopillar array) as the gate insulator, and 20-nm-thick amorphous $WO_3$/metal bilayer as the gate electrode. The channel length and width were 800 µm and 400 µm, respectively. (**b**) Cross-sectional TEM image of the device. The layer structure is clearly seen. (**c**) $R_s$–$T$ curves of the $SrCoO_x$ layer: (A) virgin state, (B) after applying a negative $V_g$ of −3 V, (C) subsequent application of +3 V (dotted line). The inset shows the cyclability at RT in air. Reversible switching of $R_s$ (ON/OFF ratio ~10$^3$) is realized. (**d**) $m$–$T$ curves of the $SrCoO_x$ layer at states A–C in (**c**), measured under $H$ = 20 Oe applied parallel to the in-plane direction. The inset shows a magnetic hysteresis loop at 10 K at states A and B. The present device is reversibly operable from AFM-insulator $SrCoO_{2.5}$ (brownmillerite structure) to FM-metal $SrCoO_3$ (perovskite structure) by applying a voltage of ±3 V for 2–3 s.

**Figure 4.** Switching mechanism of the electromagnetic device. (**a**, **b**) Retention-time ($t$) dependence of gate current, $I_g$, (top) and sheet resistance, $R_s$, (bottom) for the $SrCoO_x$ layer under application of various ±$V_g$'s from 1.5 to 3.0 V at intervals of 1 s, where a negative $V_g$ was initially applied for oxidation (**a**) and a positive $V_g$ was subsequently applied for reduction (**b**). $I_g$ and $R_s$ were respectively measured during and after $V_g$ application. The dotted lines indicate the $R_s$ value of the virgin $SrCoO_{2.5}$ layer. (**c**) Electron-density dependence of $R_s$ under application of various $V_g$'s. The electron density was calculated as the integrated value of the $I_g$–$t$ plots in (**a**, **b**). The ideal electron density required for the redox reaction between $SrCoO_{2.5}$ and $SrCoO_3$ is 7.0×10$^{16}$ cm$^{-2}$. The universal change in $R_s$, regardless of the applied $V_g$, indicates that all the provided electrons were used for the electrochemical redox reaction of the $SrCoO_x$ layer, obeying Faraday's laws of electrolysis, and that the device operation can be controlled by current density.

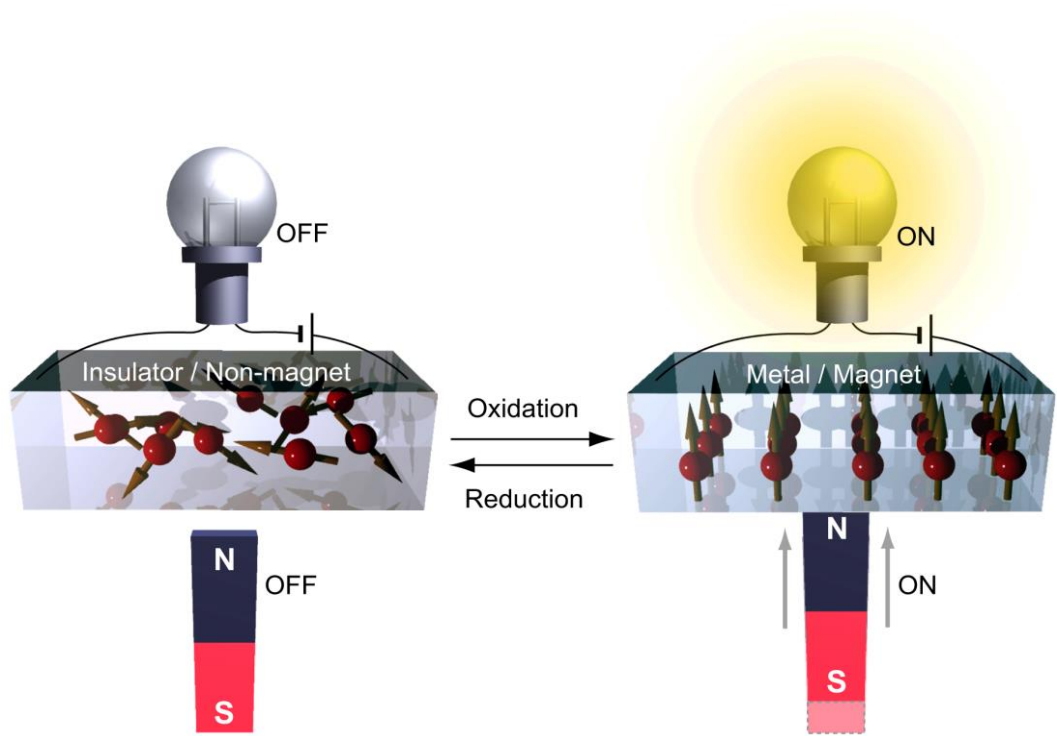

**Figure 1.**

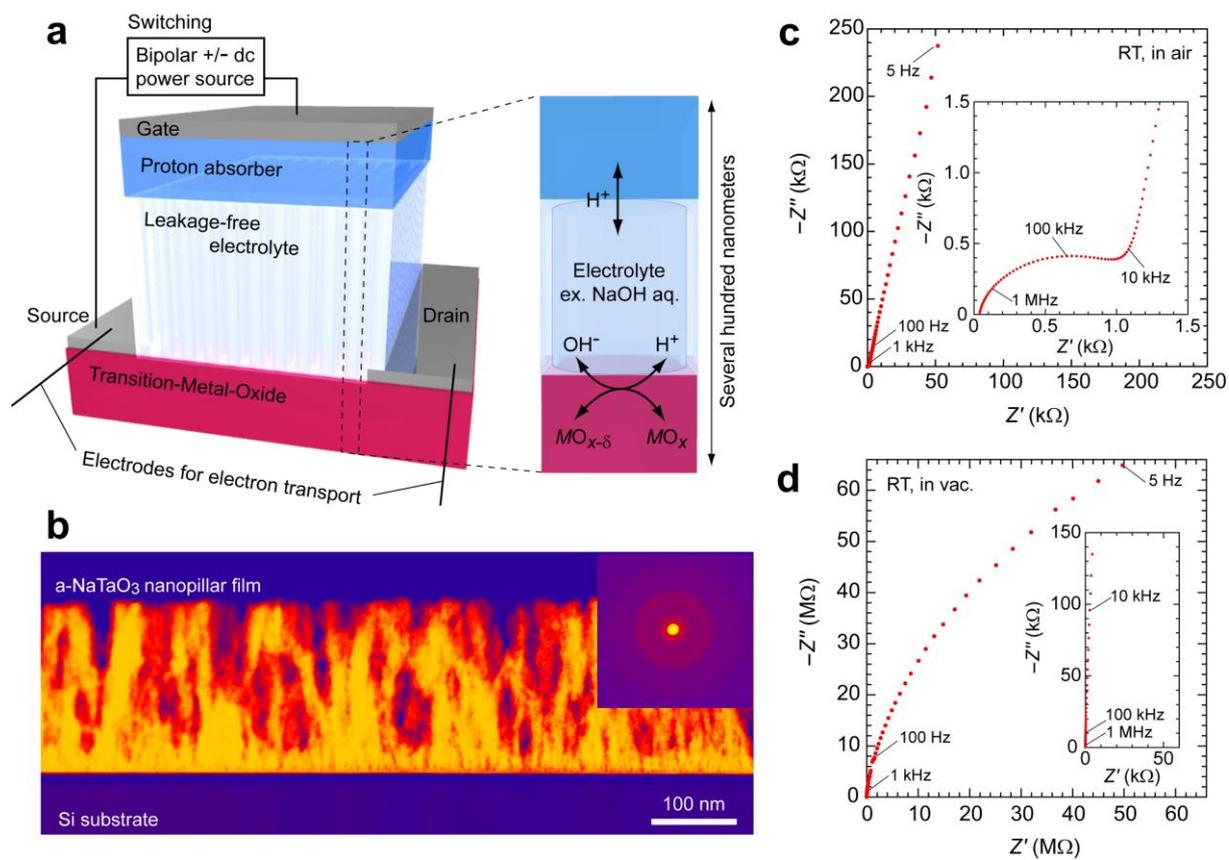

**Figure 2.**

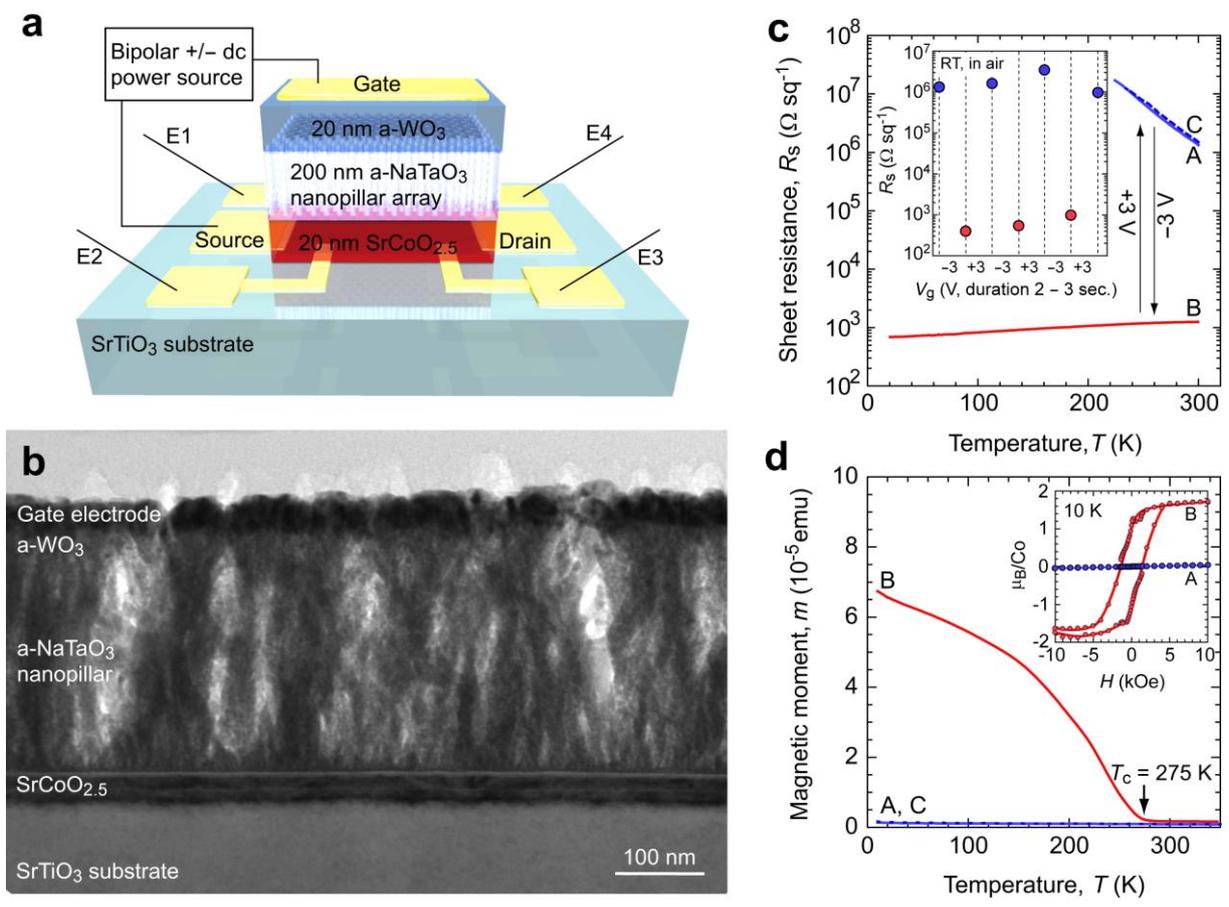

**Figure 3.**

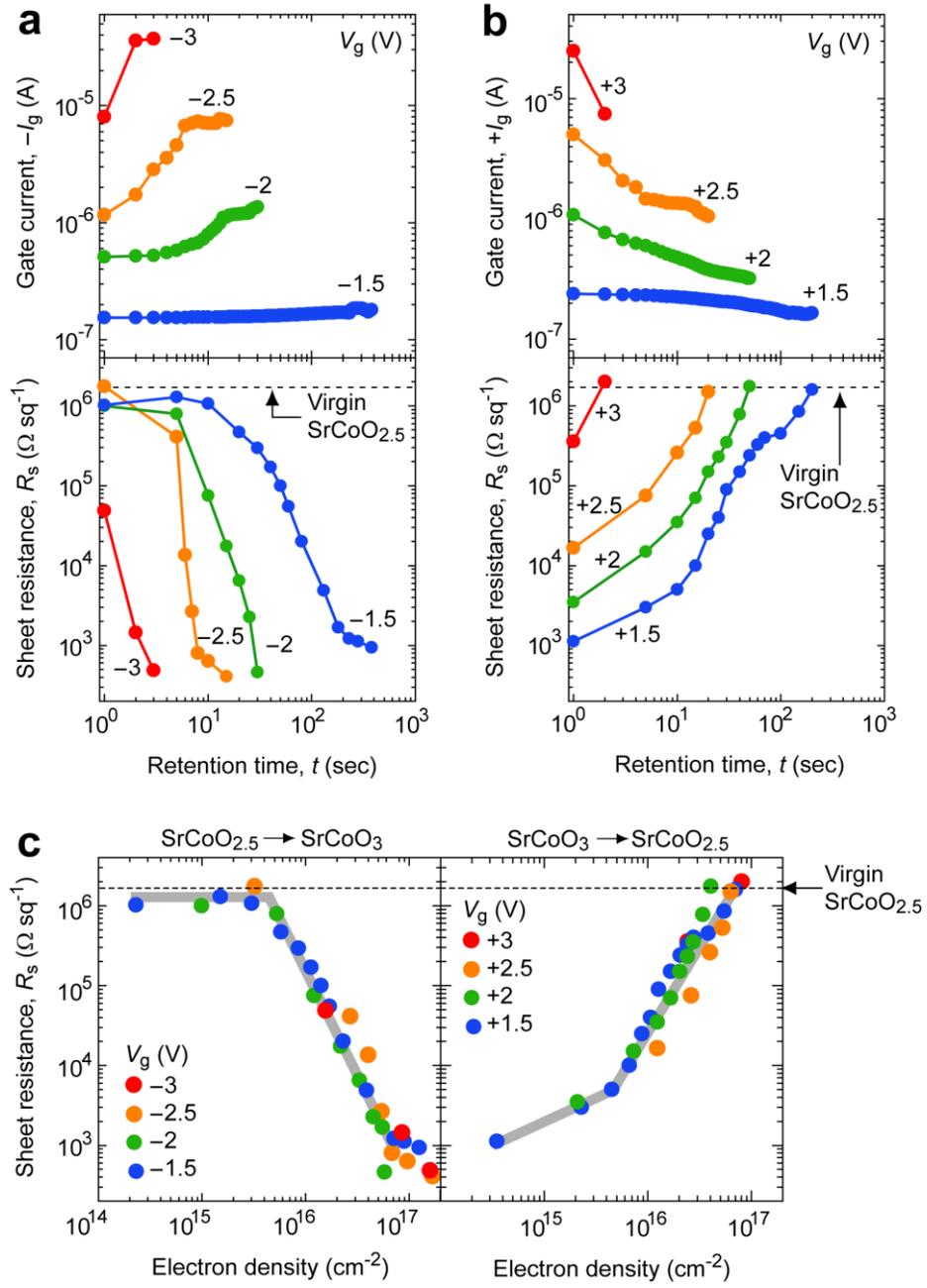

**Figure 4.**